\documentclass[11pt]{article}

\usepackage[a4paper,margin=2.4cm]{geometry}
\usepackage[utf8]{inputenc}
\usepackage[T1]{fontenc}
\usepackage{amsmath,amssymb}
\usepackage{booktabs}
\usepackage{multirow}
\usepackage{array}
\usepackage{xcolor}
\usepackage{graphicx}
\usepackage{caption}
\usepackage{enumitem}
\usepackage{url}
\usepackage[colorlinks=true,linkcolor=blue!50!black,citecolor=green!40!black,urlcolor=blue!50!black]{hyperref}
\usepackage{tikz}
\usetikzlibrary{arrows.meta,positioning,shapes.geometric,fit}

\newcommand{\scitaste}{SciTaste-Bench}
\newcommand{\dimagent}{\textsc{DimAgent-v3}}
\definecolor{dimgray}{gray}{0.55}

\newcommand{\papertitleplain}{The Scientific Research Taste Benchmark (SciTaste-Bench): A Multi-Disciplinary Benchmark for Assessing Large Language Models' Judgment of Research Taste in Scientific Papers}
\newcommand{\papertitledisplay}{\textbf{The Scientific Research Taste Benchmark (\scitaste):}\\
A Multi-Disciplinary Benchmark for Assessing Large Language Models'\\
Judgment of Research Taste in Scientific Papers}

\title{%
\texorpdfstring{%
\noindent\makebox[\textwidth]{%
\raisebox{-0.06cm}{\includegraphics[height=1.0cm]{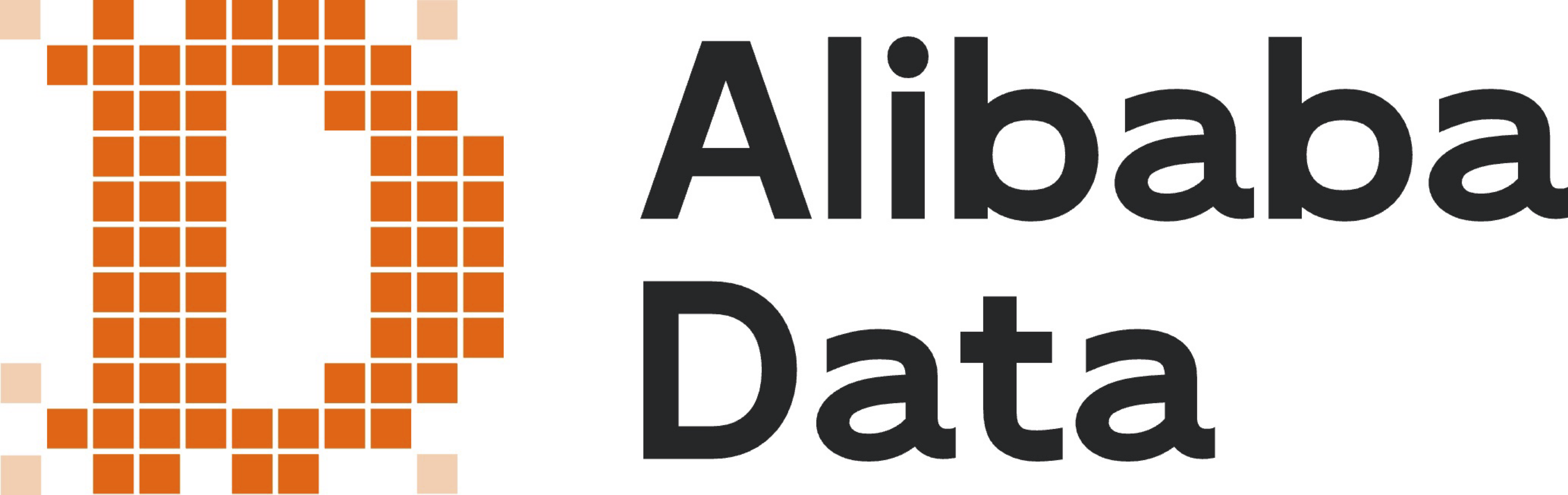}}%
\hfill
\raisebox{-0.03cm}{\includegraphics[height=1.0cm]{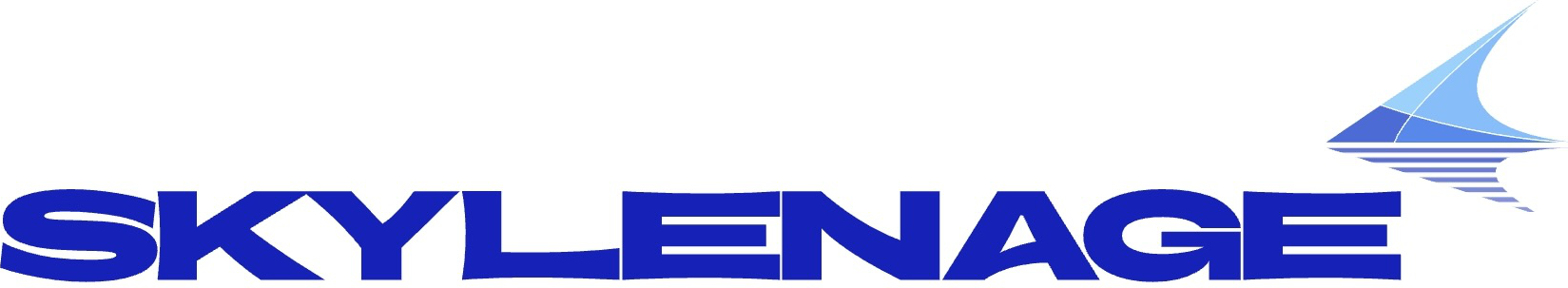}}%
}\\[-0.2em]
\rule{\textwidth}{0.5pt}\\[0.7em]
\papertitledisplay
}{\papertitleplain}%
}

\author{Xinyi Gao\textsuperscript{1,$\dagger$}, Xiaogang Li\textsuperscript{2,$\dagger$}, Jialu Nie\textsuperscript{2,$\dagger$},
Yuting Fu\textsuperscript{2}, Jingye Xu\textsuperscript{2}, Xinyan Liu\textsuperscript{2},\\
Muxi Zheng\textsuperscript{3}, Xiao Zhou\textsuperscript{2}, Wenbo Li\textsuperscript{2},
Xuan Ren\textsuperscript{2}, Hu Wei\textsuperscript{2,*}, Bing Zhao\textsuperscript{2,*}\\
\textsuperscript{1}University of Chinese Academy of Sciences\\
\textsuperscript{2}Alibaba Group\\
\textsuperscript{3}Tsinghua University\\
\texttt{kongwang@alibaba-inc.com}\quad \texttt{xiongdao@alibaba-inc.com}}

\date{\today}

\makeatletter
\renewcommand{\maketitle}{%
  \vspace*{-1.0cm}
  \begin{center}
    {\LARGE\bfseries \@title \par}
    \vspace{0.9em}
    {\large \lineskip .5em \@author \par}
    \vspace{0.55em}
    {\normalsize \@date}
  \end{center}
  \vspace{0.25em}
}
\makeatother

\begin{document}
\maketitle
{\let\thefootnote\relax\footnotetext{%
\textsuperscript{$\dagger$}These authors contributed equally to this work.\quad
\textsuperscript{*}Corresponding authors.}}

\begin{abstract}

Whether large language models can assess the innovativeness of academic
publications is important in literature reviews, research funding screening,
and academic recommendations. We introduce the \textbf{Scientific Research Taste Benchmark (\scitaste)}, a gold-standard benchmark covering \textbf{six
disciplines}: physics, mathematics, chemistry, biology, medicine, and finance. \scitaste{} measures research taste, that is, the
ability to judge the scientific value and potential of a paper, of which
innovativeness is only one dimension among the seven assessed. \scitaste{}
integrates metadata from NASA ADS, OpenAlex, and arXiv, aligns papers
to field-specific classification systems, and evaluates each paper using the
seven-dimensional \emph{Academic Taste Index}. The weights for
each dimension are derived from a bi-objective machine-learning derivation (five regressors, including random forests) and are
\emph{shared across all six disciplines}. This benchmark includes \textbf{300
selected papers (50 per discipline)} and a backup pool of \textbf{5{,}700
papers} (950 papers per field) covering 1996--2023. To allow external knowledge
to influence model decisions, \scitaste{} supports four evaluation modes:
\texttt{pure\_qa}, \texttt{native\_search}, our self-built structured tool-calling agent
\dimagent{} (9 bibliometric tools in ADS/OpenAlex), and
\texttt{dimagent\_v3\_native}. Model outputs are evaluated with two
complementary schemes: one combines correlation with dimension-level accuracy,
and the other is based on the WAPE relative error. In the evaluation of ten
state-of-the-art large language models, the tool-based structured evaluation
generally outperformed the toolless and raw web search modes, but the
improvement was model-dependent. Physics is the easiest subject, while
chemistry, biology, and medicine are still more challenging. By measuring how
well large language models judge the value potential of academic work across
fields, \scitaste{} supplies the value-judgment substrate that auto-research
systems need to screen, rank, and prioritize candidate ideas.
We release the dataset, the evaluation harness, and the scoring code to support
reproducible research on scientific-innovation assessment; all data and code are
publicly available on Hugging Face at \url{https://huggingface.co/datasets/skylenage/SciTaste-benchmark}.

\end{abstract}

\vspace{0.6em}
{\small\noindent\textbf{Keywords:} large language models; research-paper
innovativeness; scientific benchmark; scientometrics; multi-disciplinary
evaluation\par}

\section{Introduction}
\label{sec:intro}

The evaluation of the innovativeness of research papers is located at the
intersection of scientometrics and natural language processing. Citation-based
metrics have a long history~\cite{funk2017dynamic,park2023papers} , and recent
studies have shown that large language models are already being utilized to
produce scientific feedback \cite{liang2024can} , generate and rank research
ideas \cite{si2025can} , and act as reviewers or judges
\cite{zheng2023judging,nguyen2026peerreview} . However, most existing
evaluations focus on a single discipline, often physics or NLP, and tend to use
small or loosely structured samples. Furthermore, they rarely control whether
the model has access to external knowledge. This leaves three questions open:

\begin{enumerate}[leftmargin=*]
  \item Does a model's ability to judge innovativeness vary significantly across
        disciplines?
  \item How does the model's performance change when it is equipped with retrieval
        tools, native web search, or structured bibliometric database queries?
  \item Do models show systematic bias in data-sparse or sensitive domains such as
        biomedicine and the social sciences?
\end{enumerate}

\scitaste{} is designed to address these questions. The main contributions are:

\begin{enumerate}[leftmargin=*]
  \item \textbf{A unified and reproducible construction pipeline} integrates multi-
        source metadata, classifies papers according to discipline-specific taxonomies,
        and scores them using a seven-dimensional Academic Taste Index.
        The index weights are derived by machine learning and are uniform across all
        six disciplines (Sec.~\ref{sec:physics}).
  \item \textbf{Cross-discipline comparability under one fixed weight vector} covers
        the fields of physics, mathematics, chemistry, biology, medicine, and finance,
        producing 300 selected papers and 5{,}700 backup papers (950 per discipline)
        with unified metadata (Sec.~\ref{sec:extend}).
  \item \textbf{\dimagent{}, a self-built structured tool-calling agent} exposes nine
        bibliometric tools over NASA ADS and OpenAlex, allowing the isolation and
        analysis of the effect of structured external knowledge (Sec.~\ref{sec:agent}).
  \item \textbf{A dual scoring protocol} combines a correlation/dimension composite
        with WAPE relative-error grades, alongside an empirical study of ten LLMs
        across four modes (Sec.~\ref{sec:scoring}, \ref{sec:experiments}).
  \item \textbf{A foundation for auto-research.} Because \scitaste{} provides fixed,
        gold-standard judgments of scientific value across six disciplines, it can
        serve as the value-prioritization substrate for autonomous research systems,
        laying groundwork for future auto-research in science.
\end{enumerate}

\section{Related Work}
\label{sec:related}

\paragraph{Scientometric measures of impact and disruption.}

Citation dynamics has a heavy-tailed distribution, often described as a
discretised lognormal or a hooked power-law distribution depending on the field
and age \cite{thelwall2016distributions}. Beyond simple citation counts, the CD
index and its variants measure the extent to which a paper consolidates versus
disrupts existing knowledge \cite{funk2017dynamic}. Large-scale analysis shows
that papers and patents become less disruptive over time~\cite{park2023papers}.

\paragraph{LLMs for scientific evaluation.}

Large-scale language models can generate plausible peer review feedback
\cite{liang2024can} and novel research ideas \cite{si2025can}, and LLM-as-a-
judge paradigms are now common in open-ended evaluation
\cite{zheng2023judging}. A growing literature surveys LLM-based peer review and
its reliability challenges \cite{nguyen2026peerreview}. This study complements
this line of research by providing a structured, multi-discipline gold-standard
set with fixed ground-truth scores rather than free-text review judgments.

\paragraph{Tool-augmented LLMs.}

Tool calls targeting a large number of real-world APIs have been studied at
scale \cite{qin2024toolllm}. \dimagent{} is an example of this idea applied to
bibliometrics. The method does not rely on open web searches, but instead uses
nine deterministic tools to retrieve raw metadata from curated academic
databases. It must then reason from that evidence to produce a score.

\section{Construction of the Six-Discipline Evaluation Set}
\label{sec:physics}

\subsection{Overall architecture}

\begin{figure}[t]
\centering
\resizebox{\columnwidth}{!}{%
\begin{tikzpicture}[
  node distance=0.9cm and 1.1cm,
  box/.style={draw,rounded corners=2pt,align=center,minimum height=0.9cm,minimum width=2.6cm,font=\small},
  src/.style={box,fill=blue!8},
  proc/.style={box,fill=orange!10},
  dst/.style={box,fill=green!10},
  arr/.style={-{Latex[length=2.2mm]},thick,dimgray}
]
\node[src] (ads) {NASA ADS API};
\node[src,right=of ads] (oa) {OpenAlex API};
\node[src,right=of oa] (arx) {arXiv API};
\node[proc,below=0.9cm of oa] (crawl) {multi-source fusion};
\node[proc,below left=0.9cm and -0.6cm of crawl] (cls) {discipline classification};
\node[proc,below right=0.9cm and -0.6cm of crawl] (met) {seven-dimension scoring};
\node[proc,below=1.0cm of met] (qc) {quality split};
\node[dst,below left=0.9cm and -0.8cm of qc] (g) {selected 300};
\node[dst,below right=0.9cm and -0.8cm of qc] (o) {backup 5{,}700};
\draw[arr] (ads.south) to[out=-90,in=135] (crawl.north west);
\draw[arr] (oa.south) -- (crawl.north);
\draw[arr] (arx.south) to[out=-90,in=45] (crawl.north east);
\draw[arr] (crawl.south) to[out=-90,in=60] (cls.north);
\draw[arr] (crawl.south) to[out=-90,in=120] (met.north);
\draw[arr] (cls.south) to[out=-90,in=180] (qc.west);
\draw[arr] (met.south) -- (qc.north);
\draw[arr] (qc.south) to[out=-90,in=60] (g.north);
\draw[arr] (qc.south) to[out=-90,in=120] (o.north);
\end{tikzpicture}}
\caption{The data collection and routing architecture for the evaluation set fuses three sources, classifies and scores them on seven dimensions, and then splits the data into the selected evaluation set and the backup evaluation set.}
\label{fig:pipeline}
\end{figure}
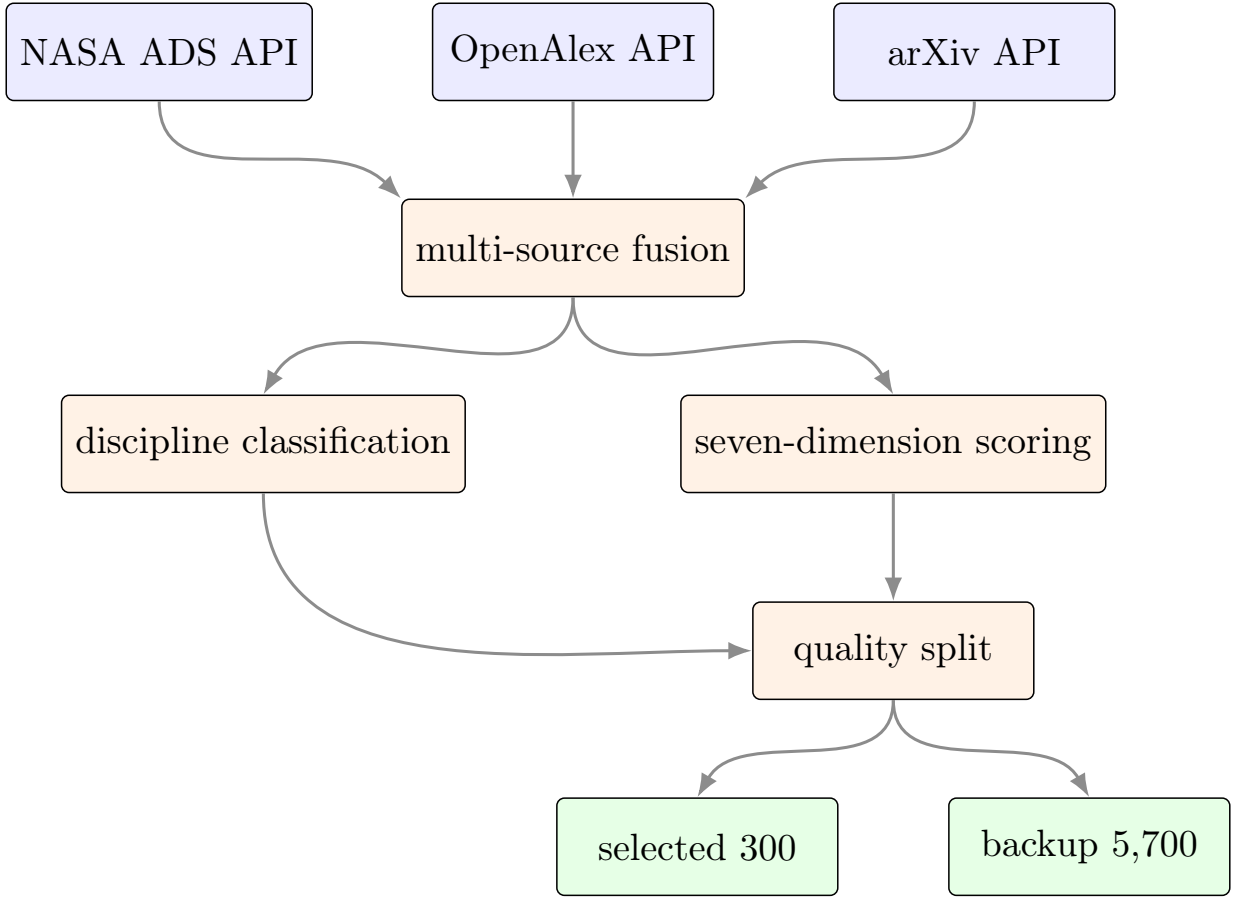

Figure~\ref{fig:pipeline} illustrates the data processing pipeline. The
indexing layer collects records from three sources. The fusion layer resolves
conflicts, the classifier assigns discipline and sub-discipline labels, and the
index engine computes the seven-dimensional index. The final quality checker
separates papers into the selected evaluation set (300 papers) and the backup
evaluation set (5{,}700 papers).

\subsection{Multi-source data collection}

Three sources feed the fusion layer, each with a distinct role across
disciplines. NASA ADS serves as the backbone for physics, OpenAlex is the
backbone for the five non-physics disciplines (Sec.~\ref{sec:extend}), and
arXiv supplies preprint metadata.

\paragraph{NASA ADS.}

ADS is the authoritative database for physics literature
\cite{accomazzi2024decades}. It contains detailed information on all related
fields, such as \texttt{bibcode}, \texttt{title}, \texttt{abstract},
\texttt{doi}, \texttt{bibstem}, \texttt{year}, \texttt{citation\_count},
\texttt{authors}, \texttt{references}, and \texttt{orcid\_pub}.

\paragraph{OpenAlex.}

OpenAlex supplies metadata that ADS may lack \cite{priem2022openalex}. For the
dimension of disruptiveness, the calculation of the CD index uses the
platform's full citing and referenced-work lists. For the dimension of
conceptual innovation, its Concepts/Topics taxonomy is utilized, and for the
dimension of team collaboration, authorship records containing institution and
country information are used.

\paragraph{arXiv (preprints).}

The arXiv XML API provides preprint metadata, and its subject-class field
serves as the highest-priority signal for discipline classification.

\paragraph{Conflict resolution.}

When the three sources disagree, we apply fixed precedence rules. Title,
publication year, and citation count follow ADS. Abstracts follow the
precedence ADS $>$ OpenAlex $>$ arXiv, and discipline labels follow arXiv $>$
journal inference $>$ keyword matching $>$ title matching.

\subsection{Discipline classification}
\label{sec:classify}

Each academic discipline is organized according to a second-level sub-
discipline taxonomy (Table~\ref{tab:tax}). This taxonomy is aligned with the
graduate-education discipline catalogue while being granular enough to provide
each sub-discipline with a roughly balanced pool of papers. Because physics and
the other five disciplines rely on different backbone databases, they use two
different labeling mechanisms.

\paragraph{Physics: a four-stage cascade over ADS/arXiv metadata.}

Eight categories are defined in physics: classical physics, condensed matter
physics, quantum physics, applied physics, high energy physics, computational
physics, interdisciplinary physics and astrophysics. A paper's label is
determined by a four-stage cascade (Table~\ref{tab:classify}). arXiv subject
classification is used where possible. If this is not available, the
publication venue is matched against discipline-characteristic journals. If a
decision still cannot be reached, controlled keywords are matched in the title
and abstract; the final fallback is title matching. A paper receives the label
from the first stage that fires. For example, \texttt{quant-ph} maps directly
to quantum physics. At the journal stage, bibstems such as PRB or NATMAT
indicate condensed matter, whereas APJ or MNRAS indicate astrophysics. At the
keyword stage, two or more controlled-vocabulary hits, such as
\emph{superconductivity} and \emph{graphene} for condensed matter, determine
the label. Papers that fail all four stages remain unclassified and are routed
to the backup evaluation set.

\paragraph{The other five disciplines: taxonomy-driven collection via
OpenAlex.}

The fields of mathematics, chemistry, biology, medicine, and finance do not
have a system equivalent to arXiv's subject classes. Therefore, the labels for
these fields are assigned \emph{by construction} rather than post hoc. Each
sub-discipline in Table~\ref{tab:tax} is anchored to an OpenAlex subfield
identifier and a set of controlled keywords that describe the area. Candidate
papers from OpenAlex are collected by selecting article-type records whose
primary topic belongs to the relevant subfield and whose text matches the
keywords, sorted by citation count in descending order. If the keyword search
returns too few records, the subfield filter alone is used to back-fill the
candidate pool. After cross-sub-discipline deduplication, where each paper
belongs to only one sub-discipline, roughly 125 papers per sub-discipline are
retained through stratified sampling across composite-score quartiles. Each
paper is labeled by the collection cell from which it was drawn. This design
avoids between-sub-discipline misclassification, while the subfield filter and
keyword query preserve topical purity within each cell. The result is eight
sub-disciplines for mathematics, chemistry, biology, and finance, and seven for
medicine (Table~\ref{tab:tax}).

\begin{table}[h]
\centering\small
\caption{Four-stage cascade for physics sub-discipline classification; the first stage that fires decides.}
\label{tab:classify}
\begin{tabular}{cll}
\toprule
Priority & Basis & Rule \\
\midrule
1 & arXiv category    & direct mapping of the subject class \\
2 & journal inference & keyword match on the publication venue \\
3 & keyword matching  & $\ge 2$ controlled-keyword hits in title/abstract \\
4 & title matching    & $\ge 2$ keyword hits in the title \\
\bottomrule
\end{tabular}
\end{table}

\begin{table}[t]
\centering\small
\caption{Second-level sub-discipline taxonomy of the six disciplines in \scitaste{}.}
\label{tab:tax}
\begin{tabular}{@{}lp{11.4cm}@{}}
\toprule
Discipline & Second-level sub-disciplines \\
\midrule
Physics (8) & classical physics; condensed matter; quantum physics; applied
physics; high-energy physics; computational physics; interdisciplinary
physics; astrophysics \\
Mathematics (8) & algebra and number theory; analysis and differential
equations; geometry and topology; probability and statistics; computational
mathematics; operations research and optimization; applied mathematics; logic
and discrete mathematics \\
Chemistry (8) & organic chemistry; inorganic chemistry; physical chemistry;
analytical chemistry; materials chemistry; computational and theoretical
chemistry; chemical biology; polymers and soft matter \\
Biology (8) & molecular biology; cell biology; genetics and genomics; ecology
and biodiversity; evolutionary biology; microbiology and infection;
structural biology and biophysics; synthetic biology and bioengineering \\
Medicine (7) & clinical medicine; basic medicine; public health and
epidemiology; pharmacy and drug development; imaging medicine and
intervention; oncology; neuroscience \\
Finance (8) & asset pricing; corporate finance; market microstructure; risk
management; financial econometrics; behavioral finance; derivatives and
quantitative finance; macro-finance and financial technology \\
\bottomrule
\end{tabular}
\end{table}

\subsection{Seven-dimensional Academic Taste Index}
\label{sec:index}

The core scoring engine assigns each paper a $0$--$100$ innovation score as a
weighted sum over seven interpretable dimensions.

\begin{equation}
\text{innovation\_score}=\sum_{i=1}^{7} W_i \cdot \text{DimensionScore}_i,
\qquad \sum_i W_i = 1.0,
\end{equation}

where each $\text{DimensionScore}_i\in[0,100]$. The same seven dimensions and
the same weight vector are applied identically to every discipline, and this is
what makes cross-discipline comparison possible. Table~\ref{tab:dims} lists the
dimensions, what they measure, and their final weights.

The weight vector is itself data-driven. The training corpus is the
998-paper quantum-physics V5 evaluation dataset from an earlier
single-discipline study. Two regression analyses were run: one predicting
citation impact from the other six dimensions, and one predicting journal
impact from the other six. Each analysis trained five regressors---linear,
ridge ($\alpha{=}1.0$), lasso ($\alpha{=}0.1$), random forest (100 trees), and
gradient boosting (100 trees)---on standardized dimension scores with an
80/20 train--test split (seed 42) and five-fold cross-validation. Per-model
feature contributions (absolute standardized coefficients for linear models,
feature importances for tree ensembles) were averaged across the five
models; dimensions shared by the two analyses received the mean of their two
contributions, and the raw weights were normalized to sum to one, yielding
the final vector in Table~\ref{tab:dims}. The random forest attained a test
$R^2$ of 0.736 for citation impact but only 0.170 for journal impact,
indicating that citation behavior is largely predictable from the remaining
dimensions whereas journal placement is comparatively independent.

\begin{table}[t]
\centering\small
\caption{The seven dimensions of the Academic Taste Index and their final normalized weights.}
\label{tab:dims}
\begin{tabular}{@{}clp{6.6cm}c@{}}
\toprule
ID & Dimension & What it measures & Weight \\
\midrule
A & Citation Dynamics       & post-publication attention and impact            & 29.5\% \\
B & Knowledge Breadth       & breadth/diversity of integrated references       & 11.8\% \\
C & Semantic Novelty        & semantic distinctiveness relative to prior work  & 11.6\% \\
D & Disruptiveness          & degree of breaking with the existing literature  & 8.9\% \\
E & Conceptual Innovation   & novelty of concept/topic combinations            & 12.9\% \\
F & Team Collaboration      & internationalization and scale of the team       & 10.1\% \\
G & Journal Impact          & venue prestige (IF / JCR / CAS partition)        & 15.3\% \\
\bottomrule
\end{tabular}
\end{table}

\paragraph{Weight derivation (dual-target Random Forest).}

Which dimensions deserve the largest weights? In everyday scholarly practice,
two community-sanctioned outcomes often shape perceptions of a paper's value:
how often it is cited and where it is published. Therefore, we used citation
count and journal impact as two independent supervision targets, allowing the
data to assess how strongly each dimension influences these outcomes.
Specifically, we ran two independent Random-Forest
analyses~\cite{breiman2001random} and interpreted feature importances as
weights. Analysis 1 predicts the citation impact (A) from the six other
dimensions, and Analysis 2 predicts the journal impact (G) from the six other
dimensions. Because A and G each serve as a target in one analysis, their own
weights are taken from the \emph{other} analysis: A's weight comes from its
contribution to predicting Journal Impact (37.99\%), and G's from predicting
Citation Impact (19.73\%). Dimensions B--F use the arithmetic mean of their
importances across the two analyses. The raw weights sum to 128.86\% and are
renormalized to 100\% (Table~\ref{tab:weights}). Both forests use 100 trees
(seed 42) with an 80/20 train--test split. The fit of the citation model is
good (test $R^2=0.74$), indicating that the citation impact is largely
predictable from the other six dimensions. On the other hand, the journal model
is much weaker ($R^2=0.17$), suggesting that the journal impact has relatively
independent information. The two targets therefore provide complementary
evidence for the final weight vector.

\begin{table}[h]
\centering\small
\caption{Dual-target weight derivation. B--F average the two analyses; A and G are cross-assigned; the final column renormalizes the 128.86\% total to 100\%.}
\label{tab:weights}
\begin{tabular}{@{}lccc@{}}
\toprule
Dimension & Citation-analysis & Journal-analysis & Final (normalized) \\
\midrule
A. Citation Impact      & (target)  & 37.99\% & 29.5\% \\
B. Knowledge Breadth    & 22.98\%   & 7.52\%  & 11.8\% \\
C. Semantic Novelty     & 12.91\%   & 16.87\% & 11.6\% \\
D. Disruptiveness       & 18.14\%   & 4.69\%  & 8.9\% \\
E. Conceptual Innovation& 21.99\%   & 11.20\% & 12.9\% \\
F. Team Collaboration   & 4.24\%    & 21.73\% & 10.1\% \\
G. Journal Impact       & 19.73\%   & (target)& 15.3\% \\
\bottomrule
\end{tabular}
\end{table}

\subsection{Dimension-to-score mapping}
\label{sec:mapping}

Each dimension's raw sub-parameters are mapped onto a $0$--$100$ score.

\paragraph{A. Citation Dynamics (29.5\%).}

From \texttt{total\_citations} we blend a log-normalized score with an
empirical percentile:

\begin{equation}
s_{\text{cit}}=0.6\,\underbrace{\frac{\log(1+c)}{\log(1+c_{\max})}\times100}_{\text{log-normalized}}
+0.4\,\underbrace{\frac{\#\{c_j<c\}}{N}\times100}_{\text{percentile}} .
\end{equation}

The logarithmic term handles the heavy-tailed citation distribution
\cite{thelwall2016distributions}. The percentile term counts the fraction of
papers that are strictly below the target, anchoring the score in the empirical
distribution.

\paragraph{B--F. PCA + percentile rank.}

The set of secondary parameters is as follows. The breadth of knowledge uses
the number of references, the diversity of reference fields, the diversity of
reference journals, and the age range of references as indicators. Semantic
novelty uses nearest neighbor and mean neighborhood semantic distances, number
of preferred fields, field Shannon entropy, SPECTER distance to the reference
list, SPECTER bridge score, and reference semantic dispersion. Disruptiveness
uses the CD index and disruptive citation count. Conceptual Innovation uses
core-concept score, concept count, core-topic score, and topic span. Team
Collaboration is measured by the country count and the author count. Missing
entries are imputed first by the column mean and then by zero. The resulting
matrix is standardized and projected onto the first principal component. If PC1
is negatively correlated with the standardized variables, the sign is reversed.
The projection is then mapped to $[0,100]$ by mean-rank normalization:

\begin{equation}
r_{\text{norm}}=\frac{\operatorname{rank}(z)-1}{N_{\text{valid}}-1}\times100.
\end{equation}

\paragraph{G. Journal Impact (15.3\%).}

This score is a fixed-weight blend of the impact factor (70\%) and two
partition schemes (15\% each):

\begin{equation}
s_G=0.70\,s_{\text{IF}}+0.15\,s_{\text{JCR}}+0.15\,s_{\text{CAS}}.
\end{equation}

The coefficients 0.70/0.15/0.15 were set from expert experience and validated
through expert inspection of the final paper scores.

The impact factor score is the percentile rank of the journal's impact factor
among observed values; missing entries receive 0 points. Both partition schemes
enter the score: JCR (Journal Citation Reports) quartiles and CAS (Chinese
Academy of Sciences zones) map Q1/zone-1$\to$100, Q2/zone-2$\to$75, Q3/zone-3$\to$50, and
Q4/zone-4$\to$25, with missing partitions again scoring 0.

\subsection{Dataset record format}

Each record contains identifiers, text, year, and the ground-truth total plus
seven per-dimension scores (~\ref{lst:record}).

\begin{figure}[h]
\begin{minipage}{\textwidth}
\small
\begin{verbatim}
{
  "bibcode": "1998PhRvA..58.3983S",
  "title": "Nonadiabatic dynamics in evaporative cooling ...",
  "abstract": "Magnetically trapped neutral atoms can be cooled ...",
  "year": 1998,
  "ground_truth_score_100": 46.36,
  "ground_truth_dimensions": {
    "citation_impact": 51.55,
    "knowledge_breadth": 35.33,
    "semantic_novelty": 34.39,
    "disruptiveness": 79.22,
    "conceptual_innovation": 44.51,
    "team_collaboration": 13.77,
    "journal_impact": 57.58
  }
}
\end{verbatim}
\end{minipage}
\caption{A single \scitaste{} record (physics).}
\label{lst:record}
\end{figure}

\section{Six-Discipline Coverage and Unification}
\label{sec:extend}

\subsection{Coverage and collection}

\scitaste{} covers six academic fields: physics, mathematics, chemistry, biology,
medicine, and finance. Physics is collected through NASA ADS and organized into
eight sub-disciplines (Sec.~\ref{sec:physics}). The five non-physics
disciplines are collected and aligned to the same framework in five stages: (1)
define a second-level taxonomy aligned with the graduate-education discipline
catalogue; (2) collect candidate papers through SCI replacement and OpenAlex
retrieval, align them by title/DOI/bibcode, and assign an internal work
identifier linked to the corresponding OpenAlex record when no ADS bibcode
exists; (3) recompute the seven-dimensional scores with the unified scoring
pipeline; (4) apply stratified sampling and manual verification; and (5)
produce unified output for cross-discipline merging.

\subsection{Unification principles}

This unified set follows two principles. \textbf{Weight consistency:} all six
disciplines share the same weight vector (Table~\ref{tab:dims}), making their
scores comparable. \textbf{Field mapping:} Chinese dimension names are mapped
to stable English keys (\texttt{citation\_impact}, \texttt{journal\_impact},
etc.), raw dimension scores are scaled to $0$--$100$, and the composite score
is recomputed.

\subsection{Sampling criteria}

For each discipline, the 50 selected papers cover the major second-level sub-
disciplines, balance journal tiers Q1--Q4 under a combined CAS-2025 and JCR-2025 tier map, emphasize 2000--2020 papers
with relatively stable citation data, and span the score range so that the set
retains discriminative power. Each sample is then manually checked for title,
abstract, discipline, and journal consistency.

\subsection{Dataset statistics}

Table~\ref{tab:stats} summarizes \scitaste{}. The released data set consists of a
\emph{selected} split (300 papers, 50 per discipline) and a \emph{backup} split
(5{,}700 papers, 950 per discipline) that supports refresh and re-sampling.

\begin{table}[h]
\centering\small
\caption{\scitaste{} dataset statistics.}
\label{tab:stats}
\begin{tabular}{@{}ll@{}}
\toprule
Metric & Value \\
\midrule
Number of disciplines            & 6 \\
Selected papers                  & 300 (50 per discipline) \\
Backup papers                    & 5{,}700 (950 per discipline) \\
Publication-year range           & 1996--2023 \\
Mean composite score             & 53.33 \\
Journal-tier balance (Q1/Q2/Q3/Q4) & 78 / 73 / 76 / 73 \\
\bottomrule
\end{tabular}
\end{table}

Each selected record is additionally assigned \texttt{discipline},
\texttt{subcategory\_id}/\texttt{name}, \texttt{journal\_tier},
\texttt{journal\_name}, and \texttt{openalex\_id} to ensure reproducibility.

\section{Tool-Calling Agent \dimagent{}}
\label{sec:agent}

\subsection{Design}

In agent mode, the model does not evaluate papers based solely on parametric
memory. Instead, it issues \emph{tool calls} to nine special tools.
\dimagent{} is not an off-the-shelf component: the agent framework was designed
and implemented in-house by the authors---the nine-tool interface, the two-round
call-then-score loop, and the reviewer system prompt are all our own---rather
than adapted from an existing agent codebase. Each
tool queries NASA ADS or OpenAlex for non-physics domains in real-time and
returns \emph{raw} evidence rather than precomputed results. The model must
integrate this evidence and decide the score itself based on it. Since the
dataset provides \texttt{bibcode}, a separate search step is not required.

\subsection{The nine tools}

Table~\ref{tab:tools} lists each tool, the raw data it returns, and the
corresponding dimensions. Each tool receives one \texttt{bibcode} argument.

\begin{table}[t]
\centering\small
\caption{The nine bibliometric tools exposed to \dimagent{}.}
\label{tab:tools}
\begin{tabular}{@{}lp{6.8cm}l@{}}
\toprule
Tool & Raw data returned & Supports \\
\midrule
\texttt{get\_citation\_data}      & citation count, reference count & citation\_impact \\
\texttt{get\_journal\_info}       & journal name, pub. date, DOI & journal\_impact \\
\texttt{get\_reference\_list}     & reference count + top-20 titles & knowledge\_breadth \\
\texttt{get\_concept\_data}       & keyword list (up to 15) & knowledge\_breadth (aux) \\
\texttt{get\_author\_data}        & author count, names, affiliations, countries & team\_collaboration \\
\texttt{get\_citation\_network}   & citing-paper list (up to 15) & disruptiveness (aux) \\
\texttt{get\_topic\_data}         & topic keyword list & conceptual\_innovation \\
\texttt{get\_disruptiveness\_data}& references + citation overlap for 5 citing papers & disruptiveness \\
\texttt{get\_keyword\_rarity}     & corpus frequency + rarity score per keyword & semantic\_novelty \\
\bottomrule
\end{tabular}
\end{table}

\subsection{Agent loop}

\begin{figure}[t]
\centering
\resizebox{0.86\columnwidth}{!}{%
\begin{tikzpicture}[
  box/.style={draw,rounded corners=2pt,align=center,minimum height=0.9cm,minimum width=3.2cm,font=\small},
  arr/.style={-{Latex[length=2.4mm]},thick,dimgray}
]
\node[box,fill=blue!8] (in) {Input: \texttt{bibcode}\\(from the selected evaluation set)};
\node[box,fill=orange!10,right=1.1cm of in] (r1) {Round 1:\\ parallel call of 9 tools};
\node[box,fill=gray!10,right=1.1cm of r1] (db) {NASA ADS / OpenAlex\\ raw evidence};
\node[box,fill=green!10,below=0.9cm of r1] (r2) {Round 2:\\ reasoning + scoring JSON};
\draw[arr] (in) -- (r1);
\draw[arr] (r1) -- node[above,font=\scriptsize]{query} (db);
\draw[arr] (db.south) to[out=-90,in=0] node[right,font=\scriptsize]{raw data} (r2.east);
\end{tikzpicture}}
\caption{\dimagent{} agent loop: after one round of parallel tool calls, one round of scoring is performed.}
\label{fig:agent}
\end{figure}
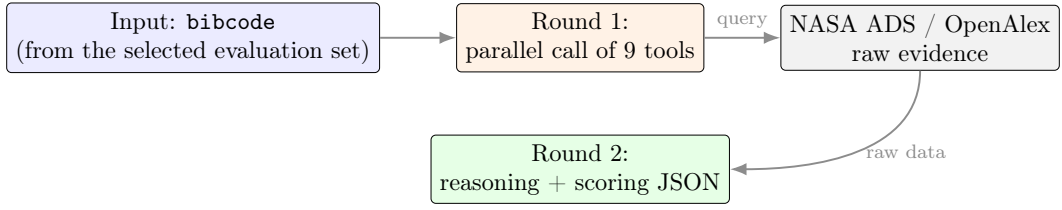

This loop is deliberately designed to be compact (see Figure~\ref{fig:agent}).
The system prompt casts the model as a senior academic reviewer and prescribes
three steps: run all nine tools in parallel, analyze the returned raw data,
then score each dimension with $0$--$100$ and calculate a weighted composite
score. Given a \texttt{bibcode}, nine simultaneous calls are sent in Round 1.
After receiving the raw data payload, Round 2 outputs an exact JSON object
containing the composite score, seven dimension scores, a brief reasoning
trace, and self-reported confidence. This design isolates structured
bibliometric data as the only external input.

\subsection{Backend routing}

Physics papers are requested directly from NASA ADS. Non-physics papers are
routed through OpenAlex: \texttt{openalex\_bibcode\_map} converts the internal
identifier to an OpenAlex work ID before calling the OpenAlex API. To comply
with OpenAlex's rate limits, we enforce a minimum request interval and include
a \texttt{mailto} field in the User-Agent to join the polite pool.

\section{Evaluation Protocol and Scoring}
\label{sec:scoring}

\subsection{Four evaluation modes}

\scitaste{} separates the data set from the evaluation harness and supports four
complementary modes (Table~\ref{tab:modes}). The two \texttt{dimagent} modes
are driven by \dimagent{}, the agent we designed and built for this benchmark
(Sec.~\ref{sec:agent}), so the tool interface and the scoring prompt are held
fixed across models. The verbatim prompt templates and
a fully worked example are given in Appendix~\ref{app:prompts}.

\begin{table}[h]
\centering\small
\caption{The four evaluation modes and the capability each probes.}
\label{tab:modes}
\begin{tabular}{@{}lp{6.0cm}p{4.6cm}@{}}
\toprule
Mode & Configuration & Capability probed \\
\midrule
\texttt{pure\_qa}            & no tools, no search; internal knowledge only & prior knowledge \& judgment \\
\texttt{native\_search}      & platform-native web search enabled & retrieval-augmented judgment \\
\texttt{dimagent\_v3}        & nine structured bibliometric tools & external-knowledge integration \\
\texttt{dimagent\_v3\_native}& nine tools + native search & structured + live retrieval \\
\bottomrule
\end{tabular}
\end{table}

\subsection{Scoring scheme 1: correlation + dimension accuracy}

This comprehensive assessment covers both an overall assessment and a detailed
assessment. In part A (50\%), the Spearman coefficient $\rho$ (15\%), Pearson
coefficient $r$ (10\%), mean absolute error and mean square error (15\%), and
$|\text{Bias}|$ (10\%) are measured. In Part B (50\%), the seven dimensions are
estimated using the weighted mean Spearman coefficient $\rho$ (30\%) and the
weighted mean absolute error (20\%), based on the weighting in
Table~\ref{tab:dims}. The final score is $\text{Part A}\times50\%+\text{Part
B}\times50\%$, and the grade classification is as follows: $\ge80$ is
excellent, $\ge60$ is moderate, $\ge40$ is poor, and $<40$ is unusable.

\subsection{Scoring scheme 2: relative error (WAPE)}

The simple mean absolute percentage error $\frac1n\sum_i |p_i-g_i|/g_i$ is
dominated by papers with low true values. We therefore adopt the weighted
absolute percentage error, following forecasting-evaluation practice
\cite{makridakis2022m5}:

\begin{equation}
\text{WAPE}=\frac{\sum_i |p_i-g_i|}{\sum_i g_i}.
\end{equation}

WAPE aggregates the total error using the true values as weights, preventing
poorly scored papers from dominating the overall results. Based on the grade
interval, this maps to a $0$--$100$ score, and the grades are A ($<5\%$), B
($5$--$10\%$), C ($10$--$20\%$), D ($20$--$30\%$), E ($30$--$50\%$), and F
($\ge50\%$) (e.g., $<5\%\to90$--$100$).

\section{Experiments and Results}
\label{sec:experiments}

\subsection{Setup}

We evaluated ten frontier large-scale language models (qwen-3.8-max,
qwen3.7-plus, qwen3.7-max, claude-fable-5, claude-opus-4-8, gpt-5.6-sol,
gemini-3.5-flash, kimi-k3, Doubao-Seed-2.1-Pro, and DeepSeek-V4-Pro) through a
unified API gateway. Where reasoning or thinking was supported, we enabled the
setting: high for the Qwen and GPT models, adaptive for the Claude models, and
standard for Gemini, Kimi, Doubao, and DeepSeek. The main experiments
were conducted under all four conditions using a dataset of 300 articles from
six academic disciplines. For each model and mode combination, the percentage
of successfully scored papers was recorded.

\subsection{Main results}

Table~\ref{tab:main} reports the Scheme-1 and Scheme-2 scores for all ten
models. gpt-5.6-sol was fully re-run after a parameter-conflict fix
(Sec.~\ref{sec:gpt}), and gemini-3.5-flash is newly added in this round.
Structured tool grounding (\texttt{dimagent\_v3}) remains the strongest single
intervention. For the three Qwen models and claude-fable-5, Scheme-2 scores
rise from roughly $20$--$33$ in the tool-free modes to $47$--$52$ (grade E
$\to$ D). claude-opus-4-8 improves only to $34.5$ in the pure tool mode, still
grade E, with the combined mode becoming its best configuration ($40.7$). There
are two exceptions. In the \texttt{dimagent\_v3} run, gpt-5.6-sol's Scheme-2
score collapses to $10.6$ (grade F) but recovers to $28.7$ (E) when native
search is added. gemini-3.5-flash is the weakest overall, scoring grade F in
all four modes, and structured tools yield only a marginal gain. The three
newly added models refine this picture. kimi-k3 stays at grade E in all four
modes, with structured tools lifting its Scheme-2 score from $25.0$ to $33.4$.
Doubao-Seed-2.1-Pro is the strongest of the newcomers, reaching grade D
($44.0$) in the combined mode, which is also its best configuration.
DeepSeek-V4-Pro shows the weakest sensitivity to external grounding: its
tool-free \texttt{pure\_qa} score ($30.2$) is already its best, and structured
tools add little ($34.5$), indicating that its parametric prior is
comparatively well matched to the task while bibliometric grounding fails to
compound.

\begin{table}[t]
\centering\small
\caption{Main results on the six-discipline 300-paper set. The Scheme-2 grade is given in brackets; superscripts mark run-level issues described below.}
\label{tab:main}
\setlength{\tabcolsep}{4pt}
\begin{tabular}{@{}llcccc@{}}
\toprule
Model & Mode & Scheme 1 & Scheme 2 & Grade & Success \\
\midrule
\multirow{4}{*}{qwen-3.8-max}
 & pure\_qa             & 21.10 & 28.10 & E & 300/300 \\
 & native\_search       & 23.20 & 31.60 & E & 300/300 \\
 & dimagent\_v3         & 28.00 & 51.80 & D & 300/300 \\
 & dimagent\_v3\_native & 27.10 & 50.30 & D & 300/300 \\
\midrule
\multirow{4}{*}{qwen3.7-plus}
 & pure\_qa             & 19.40 & 23.20 & E & 300/300 \\
 & native\_search       & 19.80 & 20.80 & E & 300/300 \\
 & dimagent\_v3         & 27.80 & 47.60 & D & 300/300 \\
 & dimagent\_v3\_native & 27.70 & 48.10 & D & 300/300 \\
\midrule
\multirow{4}{*}{qwen3.7-max}
 & pure\_qa             & 19.30 & 20.10 & E & 300/300 \\
 & native\_search       & 20.20 & 21.10 & E & 300/300 \\
 & dimagent\_v3         & 27.60 & 47.70 & D & 300/300 \\
 & dimagent\_v3\_native & 27.40 & 47.20 & D & 300/300 \\
\midrule
\multirow{4}{*}{claude-fable-5}
 & pure\_qa             & 22.30 & 27.60 & E & 269/300$^{\dagger}$ \\
 & native\_search       & 24.60 & 32.70 & E & 280/300$^{\dagger}$ \\
 & dimagent\_v3         & 26.80 & 47.00 & D & 289/300$^{\dagger}$ \\
 & dimagent\_v3\_native & 25.20 & 39.30 & E & 291/300$^{\dagger}$ \\
\midrule
\multirow{4}{*}{claude-opus-4-8}
 & pure\_qa             & 20.20 & 25.50 & E & 300/300 \\
 & native\_search       & 21.90 & 27.00 & E & 300/300 \\
 & dimagent\_v3         & 23.70 & 34.50 & E & 300/300 \\
 & dimagent\_v3\_native & 25.20 & 40.70 & D & 300/300 \\
\midrule
\multirow{4}{*}{gpt-5.6-sol}
 & pure\_qa             & 19.50 & 19.20 & F & 300/300 \\
 & native\_search       & 21.10 & 19.50 & F & 300/300 \\
 & dimagent\_v3         & 9.50  & 10.60 & F & 300/300 \\
 & dimagent\_v3\_native & 22.20 & 28.70 & E & 300/300 \\
\midrule
\multirow{4}{*}{gemini-3.5-flash}
 & pure\_qa             & 16.90 & 16.40 & F & 300/300 \\
 & native\_search       & 16.40 & 16.50 & F & 113/300$^{\ddagger}$ \\
 & dimagent\_v3         & 19.90 & 17.70 & F & 300/300 \\
 & dimagent\_v3\_native & 19.70 & 17.60 & F & 300/300 \\
\midrule
\multirow{4}{*}{kimi-k3}
 & pure\_qa             & 21.40 & 25.00 & E & 298/300$^{\dagger}$ \\
 & native\_search       & 23.20 & 28.90 & E & 300/300 \\
 & dimagent\_v3         & 24.00 & 33.40 & E & 281/300$^{\dagger}$ \\
 & dimagent\_v3\_native & 24.70 & 30.10 & E & 300/300 \\
\midrule
\multirow{4}{*}{Doubao-Seed-2.1-Pro}
 & pure\_qa             & 20.70 & 28.00 & E & 300/300 \\
 & native\_search       & 22.50 & 28.60 & E & 300/300 \\
 & dimagent\_v3         & 24.40 & 39.20 & E & 300/300 \\
 & dimagent\_v3\_native & 25.50 & 44.00 & D & 300/300 \\
\midrule
\multirow{4}{*}{DeepSeek-V4-Pro}
 & pure\_qa             & 19.10 & 30.20 & E & 300/300 \\
 & native\_search       & 20.00 & 23.40 & E & 300/300 \\
 & dimagent\_v3         & 24.00 & 34.50 & E & 300/300 \\
 & dimagent\_v3\_native & 23.60 & 33.70 & E & 300/300 \\
\bottomrule
\end{tabular}
\begin{minipage}{0.94\textwidth}\footnotesize\vspace{3pt}
$^{\dagger}$\,Gateway issues: empty or non-JSON responses and HTTP 429 rate
limiting; the affected records were dropped from the metrics and were not
recoverable by re-running.
$^{\ddagger}$\,Mass failures: the 187 incomplete \texttt{native\_search} runs
span all six disciplines and never emitted a parseable scoring JSON.
Rows with fewer than 300 successes are averaged over successful records only.
\end{minipage}
\end{table}

\subsection{Cross-discipline analysis}

Table~\ref{tab:cross} breaks down the \texttt{dimagent\_v3} mode by discipline
for qwen-3.8-max. Physics is by far the easiest ($\rho{=}0.78$, WAPE $13.1\%$,
Scheme-2 $70.4$, grade C), reflecting richer parametric coverage and cleaner
ADS grounding. Finance is the strongest non-physics discipline. Medicine has
the weakest correlation ($\rho{=}0.31$), while chemistry and biology have the
largest relative error (WAPE $\approx29\%$). The cross-model breakdown
(Table~\ref{tab:crossmodel}) shows that the physics advantage is shared by
most models: the three Qwen models, claude-fable-5, and kimi-k3 all reach
Scheme-2 between $46.9$ and $70.7$ on physics. Doubao-Seed-2.1-Pro inverts this
pattern, scoring its best in finance ($58.6$) and medicine ($49.8$) and its
worst in physics ($15.8$), a reminder that the discipline gap is partly a
property of each model's training mix rather than of the benchmark. The
claude-opus-4-8 physics cell ($6.3$) is an instability of that model's tool run
($\rho{=}{-0.16}$, WAPE $74.4\%$), not a reflection of discipline difficulty.

\begin{table}[t]
\centering\small
\caption{Results by discipline for \texttt{dimagent\_v3} (qwen-3.8-max).}
\label{tab:cross}
\begin{tabular}{@{}lcccc@{}}
\toprule
Discipline & Spearman $\rho$ & MAE & WAPE & Scheme 2 \\
\midrule
Overall    & 0.5481 & 12.86 & 24.1\% & 51.80 [D] \\
Physics    & 0.7751 & 6.21  & 13.1\% & 70.40 [C] \\
Finance    & 0.6923 & 11.90 & 22.5\% & 55.00 [D] \\
Mathematics& 0.5293 & 13.03 & 23.6\% & 52.70 [D] \\
Chemistry  & 0.5697 & 15.80 & 29.2\% & 41.70 [D] \\
Biology    & 0.4599 & 15.72 & 28.8\% & 42.50 [D] \\
Medicine   & 0.3099 & 14.51 & 26.1\% & 47.80 [D] \\
\bottomrule
\end{tabular}
\end{table}

\begin{table}[t]
\centering\footnotesize
\caption{Scheme-2 scores by discipline for \texttt{dimagent\_v3} across all ten models. $^{*}$The claude-opus-4-8 physics sub-run is anomalous ($\rho{=}{-0.16}$, WAPE $74.4\%$) and reflects an instability of that model under tool mode rather than the difficulty of the discipline.}
\label{tab:crossmodel}
\setlength{\tabcolsep}{4pt}
\begin{tabular}{@{}lccccccc@{}}
\toprule
Model & Overall & Physics & Finance & Math & Chem & Bio & Med \\
\midrule
qwen-3.8-max        & 51.8 & 70.4 & 55.0 & 52.7 & 41.7 & 42.5 & 47.8 \\
qwen3.7-plus        & 47.6 & 69.4 & 57.3 & 47.9 & 37.5 & 35.7 & 44.7 \\
qwen3.7-max         & 47.7 & 70.7 & 50.4 & 46.0 & 38.3 & 38.3 & 45.7 \\
claude-fable-5      & 47.0 & 66.9 & 48.5 & 51.4 & 38.2 & 37.0 & 42.7 \\
claude-opus-4-8     & 34.5 & 6.3$^{*}$ & 60.8 & 49.0 & 37.1 & 33.1 & 42.3 \\
gpt-5.6-sol         & 10.6 & 10.2 & 13.0 & 9.5  & 10.5 & 10.2 & 13.6 \\
gemini-3.5-flash    & 17.7 & 29.1 & 16.2 & 17.7 & 15.4 & 15.8 & 18.4 \\
kimi-k3             & 33.4 & 46.9 & 35.6 & 37.5 & 29.8 & 28.8 & 24.7 \\
Doubao-Seed-2.1-Pro & 39.2 & 15.8 & 58.6 & 51.6 & 43.6 & 40.3 & 49.8 \\
DeepSeek-V4-Pro     & 34.5 & 31.1 & 38.6 & 35.8 & 32.8 & 32.5 & 35.8 \\
\bottomrule
\end{tabular}
\end{table}

\subsection{Findings}
\label{sec:findings}

\paragraph{Structured tools beat raw web search.}

Across models, \texttt{dimagent\_v3} is usually the strongest or near-strongest
mode. Native search alone barely improves \texttt{pure\_qa} and is sometimes
worse: unstructured web results mix signal with noise, which often causes the
model to alternate between the paper's own abstract and distracting external
hits. Structured tools return more precise bibliometric evidence and are better
matched to this task. However, this advantage is not universal. gpt-5.6-sol
collapses in the pure tool setting, and gemini-3.5-flash improves only
marginally (Sec.~\ref{sec:gpt}).

\paragraph{Adding native search to tools rarely helps.}

When analyzing the three Qwen models, \texttt{dimagent\_v3\_native} tracks
\texttt{dimagent\_v3} closely, with performance differences staying within
about one point on both schemes. The pattern diverges for the Claude models:
claude-fable-5 drops from 47.0 to 39.3 on Scheme 2, while claude-opus-4-8 rises
from 34.5 to 40.7, making the combined mode its best configuration. The most
extreme case is gpt-5.6-sol: native search pulls it back from a collapsed pure-
tool run (Scheme-2 10.6) to 28.7, its only grade-E configuration. For
gemini-3.5-flash, the two tool modes are essentially tied. Among the newly
added models, Doubao-Seed-2.1-Pro is the clearest beneficiary of stacking
search on tools ($39.2\to44.0$, E $\to$ D), whereas kimi-k3 ($33.4\to30.1$) and
DeepSeek-V4-Pro ($34.5\to33.7$) gain nothing from the combination.

\paragraph{Discipline gap persists even with tools.}

Physics holds a large advantage in most models and modes, while biology and
medicine remain difficult. This points to uneven training coverage and to the
limits of ADS-style grounding outside physics. The collapsed physics sub-run of
gpt-5.6-sol's combined mode is a notable exception (Sec.~\ref{sec:gpt}).

\paragraph{Systematic over- and under-estimation.}

When using the tool-free mode, a large positive bias was observed, indicating
that models tend to over-rate papers. The structured tool reduces this bias
and, in physics, can even make it slightly negative. The direction of the bias
can still reverse: in the pure tool run of gpt-5.6-sol, a strong negative bias
occurred ($-20.2$). The median per-paper relative error reached $100\%$. In
other words, it was confirmed that more than half of the papers are severely
under-rated in this mode. The mechanism behind this reversal is examined in Section~\ref{sec:gpt}. Overall, Scheme-1 scores remain at a low level
because the weight-averaged dimensional correlation (see Part B) is heavily
dependent on three challenging dimensions: conceptual\_innovation,
semantic\_novelty, and disruptiveness. The per-dimension correlation in each of
these dimensions is near zero in all evaluation modes.

\subsection{Engineering notes and the gpt-5.6-sol case}
\label{sec:gpt}

Several practical issues arose. \textbf{Gateway issues:} the gateway
sometimes returned empty or non-JSON responses, or rate-limited requests until
the retry budget was exhausted (HTTP 429), as for claude-fable-5 on a subset of
biology, medicine, and chemistry titles and for kimi-k3 under
\texttt{dimagent\_v3}. Such records were dropped from the metrics rather than
scored as zero, so the affected success counts remain under 300; they are
reported separately rather than treated as a model weakness. \textbf{Mass failures:} gemini-3.5-flash completed only 113/300
items under \texttt{native\_search}, with failures spread across all six
disciplines; in those runs the model never emitted a parseable scoring JSON. Its other three modes completed
300/300. \textbf{Parameter compatibility:} some models reject certain
temperature, reasoning, and tool combinations, which the client handles per-
model. Claude thinking also required a larger \texttt{max\_tokens} value to
avoid truncation. \textbf{Resumption:} the harness supports \texttt{--resume}
keyed on \texttt{bibcode}, plus a retry script for transient failures.

\textbf{gpt-5.6-sol} requires a caveat; this is the same case as the strong negative bias reported in Section~\ref{sec:findings}. The early tool-calling runs for this
model were affected by a parameter-conflict bug. Specifically, combining
\texttt{temperature} and \texttt{reasoning\_effort} with tools triggered API
errors. To resolve this, these two parameters were disabled for this model's
tool calls on the client side, and a full re-run was performed. As a result,
all four modes achieved 300/300. Five transient \texttt{pure\_qa} failures
during testing were back-filled by retry. The results of this re-run are
included in Table~\ref{tab:main}. The \texttt{pure\_qa} and
\texttt{native\_search} results for this model are broadly on par with other
models, both receiving a grade of F, but \texttt{dimagent\_v3} shows systematic
severe underestimation. Scheme-2 scores were only $10.6$, with
$\rho{\approx}0.17$, bias value $-20.2$, and all disciplines received a grade
of F. Enabling native search recovered performance to $28.7$ (grade E), but the
physics sub-run destabilized ($\rho$ is undefined, Scheme-2 scores are $0.0$).
Therefore, this is not a technical issue but indicates a genuine model-level
difference. The lesson is that tool-calling configurations must be validated
per model before large runs, and a technically valid run does not guarantee
behaviorally effective tool use.

\section{Limitations and Future Work}
\label{sec:limits}

Since each discipline offers only 50 selected papers, this benchmark cannot
cover every sub-discipline in depth. Some runs on biomedical and chemistry
titles failed at the gateway---empty or non-JSON responses and HTTP 429 rate
limiting---leaving gaps that cannot always be back-filled. The dataset is English-dominant and may under-represent non-English
scholarship. Since citation counts and journal indicators drift over time, the
backup pool needs to be refreshed and re-sampled periodically. The next natural
steps would be to expand the benchmark to computer science, materials science,
earth science, and the social sciences, and to add human-expert annotations to
complement the bibliometric ground truth.

A further limitation concerns the weight vector itself. The seven dimension
weights are obtained from a single dual-target derivation trained on 998 quantum-physics papers with five regressors (including a random forest) and are
then shared across all six disciplines. We have not re-run the derivation
separately within each discipline to obtain field-specific weight vectors, so
the transferability of one common weight vector across fields remains an
assumption of the current benchmark. We regard this as one of the principal
limitations of this work; per-discipline re-derivation and comparison are left
to future work.

\section{Conclusion}
\label{sec:conclusion}

We present \scitaste{}, a six-discipline gold-standard benchmark for judging
research innovativeness. The benchmark is constructed by (i) integrating
metadata from ADS/OpenAlex/arXiv and scoring each paper with a seven-
dimensional, machine-learning-weighted Academic Taste Index under
a single weight vector shared across physics, mathematics, chemistry, biology,
medicine, and finance; (ii) designing and implementing in-house a structured
tool-calling agent, \dimagent{}, together with its nine bibliometric tools,
two-round loop, and reviewer prompt; and (iii) evaluating models with
correlation- and relative-error-based methods. Across ten large language
models, structured tool-based evaluation generally outperforms tool-free and
raw web-search modes, although this advantage is model-dependent, with
gpt-5.6-sol's collapsed pure-tool run as the main counterexample. Physics is
the easiest field, while chemistry, biology, and medicine are the most
difficult, and dimension-level evaluation remains the key bottleneck. The
dataset, the harness, and the agent---all designed and built by the
authors---are released to support reproducible research on
scientific-innovation assessment, and are available on Hugging Face at
\url{https://huggingface.co/datasets/skylenage/SciTaste-benchmark}. Beyond benchmarking, we view \scitaste{} as
infrastructure for future auto-research: autonomous systems that propose,
evaluate, and prioritize scientific ideas require exactly this ability to judge
the value potential of research papers.

\section*{Statement on AI Assistance}
Large language models were used in preparing this manuscript solely to aid and
polish the writing. A multi-hop LLM paraphrasing pipeline was applied
segment-wise to the prose for language-level editing, style polishing, and
readability, and an LLM assistant helped rephrase and reorganize author-written
sentences and paragraphs---including the exposition of the results
analysis---so that findings already established by the authors were expressed
more clearly. In every case the underlying content, analytical points, tables,
and figures were supplied by the authors. Large language models were not used to
generate or label data, design experiments, run the evaluations, derive results,
search the literature, or prove mathematical claims, and no model-generated text
was accepted without human verification. All text was reviewed, corrected, and
approved by the human authors, who take full responsibility for the content of
this paper.

\appendix
\section{Citation-count distributions}
\label{app:cite}

The distribution of citation counts is heavy-tailed. Two prevailing models are
the discretised lognormal distribution and the hooked or shifted power-law
distribution. The former is often favoured for the arts, humanities, social
sciences, engineering fields, single disciplines, and papers published $\ge3$
years ago. The latter is more commonly favoured for the natural sciences, life
sciences, recent papers, and datasets with many zero-cited items; this model
naturally supports zero citations \cite{thelwall2016distributions}. There is no
universal distribution. For cross-field comparison, percentile normalization
(e.g., the FWCI indicator) combined with non-parametric tests is preferable.
For modeling within a specific field, AIC/BIC criteria can be used to choose
between candidate distributions. For regression analysis, $\log(c+1)$ combined
with ordinary least squares is usually the most stable choice. All zero-cited
papers should be included, heterogeneity across fields should be controlled by
stratification, and goodness-of-fit should not be interpreted as evidence for a
generative mechanism.

\section{Artifact inventory}
\label{app:files}

The main outcomes published in this study include the merged dataset with its
selected and backup splits, the identifier-to-OpenAlex alignment map, the
report generator, the agent task configuration defining the nine tools and four
evaluation modes, the scoring-metrics module implementing both scoring schemes,
and the per-model run scripts, and the weight-derivation package containing
the trained regression models (including the random forest) together with their hyperparameters and
feature-importance outputs. All components are published together with the
dataset, ensuring full replication. The detailed dataset and the full code are
open-sourced on Hugging Face at \url{https://huggingface.co/datasets/skylenage/SciTaste-benchmark}.

\section{Evaluation prompts and a worked example}
\label{app:prompts}

The original prompts were written in Chinese; the blocks below are condensed
translations. All four modes share the same user template.

\begin{verbatim}
Please assess the innovativeness of the
following {discipline} paper.
Discipline: {discipline}
Title: {title}
Abstract: {abstract}
Identifier: {bibcode}
\end{verbatim}

\noindent is a mode-specific termination instruction. The system prompt of
\texttt{pure\_qa} casts the model as a senior academic reviewer, requiring it
to score the seven dimensions from its own scholarly knowledge, list the seven-
dimension rubric using the shared weight vector, and conclude with ``output the
JSON directly, no explanation''. Its user template closes with ``score the
seven dimensions from your scholarly knowledge and output JSON.''. The system
prompt of \texttt{native\_search} retains the same skeleton, but requires the
model to first retrieve the paper's citation count, venue reputation,
authorship, and field information through native web search, and to ground
every score in the retrieved data. The \dimagent{} mode uses the system prompt
summarized in Sec.~5: it consists of the nine ADS tools, the three-step
parallel-call workflow, the rule that all nine tools must be called and that
scores must be grounded in the returned payloads, and the same strict-JSON
contract (composite score, seven dimension scores, reasoning trace, and self-
reported confidence). Verbatim prompts ship with the released harness.

\paragraph{Worked example.}

Instantiating the template on a real item gives:

\begin{verbatim}
Discipline: physics
Title: Nonadiabatic dynamics in evaporative
  cooling of trapped atoms by a
  radio-frequency field
Abstract: Magnetically trapped neutral atoms can
  be cooled with the evaporation technique. ...
  [truncated]
Identifier: 1998PhRvA..58.3983S
\end{verbatim}

\noindent The model must reply with the strict JSON; the ground truth of this
item is

\begin{verbatim}
{"innovation_score": 46.36,
 "dimension_scores": {
  "citation_impact": 51.55,
  "journal_impact": 57.58,
  "conceptual_innovation": 44.51,
  "knowledge_breadth": 35.33,
  "semantic_novelty": 34.39,
  "team_collaboration": 13.77,
  "disruptiveness": 79.22}}
\end{verbatim}

\bibliographystyle{plain}
\bibliography{references}

\end{document}